\documentclass[9pt,journal,twoside,web]{ieeecolor}
\usepackage{generic}
\usepackage{cite}
\usepackage{amsmath,amssymb,amsfonts}
\usepackage{algorithmic}
\usepackage{graphicx}
\usepackage{xcolor,tabularx,array,lscape,multirow,colortbl,rotating}
\usepackage{url,bookmark,caption,subcaption,flushend}
\graphicspath{{Figures-arxiv/}} 
\usepackage{textcomp}
\def\BibTeX{{\rm B\kern-.05em{\sc i\kern-.025em b}\kern-.08em
    T\kern-.1667em\lower.7ex\hbox{E}\kern-.125emX}}
    
\begin{document}
\title{Prediction of Reaction Time and Vigilance Variability from Spatiospectral Features of Resting-State EEG in a Long Sustained Attention Task}
\author{Mastaneh Torkamani-Azar, \IEEEmembership{Student Member, IEEE}, Sumeyra Demir Kanik, \IEEEmembership{Member, IEEE}, Serap Aydin, \IEEEmembership{Member, IEEE}, and Mujdat Cetin, \IEEEmembership{Fellow, IEEE}
\thanks{This work was supported by the Scientific and Technological Research Council of Turkey (TÜBİTAK) under the grant number 116E086. \textit{(Corresponding author:
Mastaneh Torkamani-Azar.)}}
\thanks{Mastaneh Torkamani-Azar is and Sumeyra Demir Kanik was with the Signal Processing and Information Systems (SPIS) Laboratory, Faculty of Engineering and Natural Sciences, Sabanci University, Istanbul 34956, Turkey (e-mail: mastaneh@sabanciuniv.edu; sudemirkanik@gmail.com).}
\thanks{Serap Aydin is with the Department of Biophysics, Faculty of Medicine, Hacettepe University, Ankara 06100, Turkey (e-mail: serap.aydin@hacettepe.edu.tr).}
\thanks{Mujdat Cetin is with the Department of Electrical and Computer Engineering, University of Rochester, Rochester, NY 14627, USA, and with the Signal Processing and Information Systems (SPIS) Laboratory, Faculty of Engineering and Natural Sciences, Sabanci University, Istanbul 34956, Turkey (e-mail: mujdat.cetin@rochester.edu).}}

\maketitle

\begin{abstract}
Resting-state brain networks represent the intrinsic state of the brain during the majority of cognitive and sensorimotor tasks. However, no study has yet presented concise predictors of task-induced vigilance variability from spectrospatial features of the pre-task, resting-state electroencephalograms (EEG). We asked ten healthy volunteers (6 females, 4 males) aged from 22 to 45.5 to participate in 105-minute fixed-sequence-varying-duration sessions of sustained attention to response task (SART). A novel and adaptive vigilance scoring scheme was designed based on the performance and response time in consecutive trials, and demonstrated large inter-participant variability in terms of maintaining consistent tonic performance. Multiple linear regression using feature relevance analysis obtained significant predictors of the mean cumulative vigilance score (CVS), mean response time, and variabilities of these scores from the resting-state, band-power ratios of EEG signals, p\textless0.05. Single-layer neural networks trained with cross-validation also captured different associations for the beta sub-bands. Increase in the gamma (28-48 Hz) and upper beta (24-28 Hz) ratios from the left central and temporal regions predicted slower reactions and more inconsistent vigilance as explained by the increased activation of default mode network (DMN) and differences between the high- and low-attention networks at temporal regions (Brodmann’s areas 35 and 36). Higher ratios of parietal alpha (8-12 Hz) from the Brodmann's areas 18, 19, and 37 during the eyes-open states predicted slower responses but more consistent CVS and reactions associated with the superior ability in vigilance maintenance. The proposed framework and these first findings on the most stable and significant attention predictors from the intrinsic EEG power ratios can be used to model attention variations during the calibration sessions of BCI applications and vigilance monitoring systems.
\end{abstract}


\section{Introduction}
Cognitive and affective state monitoring has become a subject of interest in the development of better human-computer interfaces for patients and healthy users. Such monitoring can involve the utilization of a variety of neurophysiological biomarkers to determine the onset of fatigue and frustration, decline of motivation, lapses in attention or vigilance, onset of drowsiness and sleep spindles, or changes in the mental workload under various task difficulty levels \cite{BrouwerBook2015}. Detecting these mental states, usually through passive brain-computer interfaces (BCIs), can increase the accuracy of human-computer interactions since changes in these cognitive and affective states lead to nonstationarities in the brain electrical activity and cause challenges for automated intent inference. Inability to sustain attention in response to important but rare stimuli is one of the important factors behind lower-than-expected accuracy of BCI systems. Monitoring the attention level is, subsequently, critical in building future adaptive BCI systems and in assessing the performance of operators in monotonous and critical tasks such as the air traffic control and long-haul driving \cite{BrouwerBook2015}. 

Various methods have been introduced to estimate sustained attention during BCI execution and to adapt the classifiers or experiment environments to the detected attention level of the users \cite{Myrden2016}. In clinical settings, sustained attention is generally quantified by the number of errors and reaction delays in response to the target stimuli while the participants are attending to monotonic paradigms such as the Continuous Performance Test (CPT) or Sustained Attention to Response Task (SART). In such settings, different response styles of participants, i.e., balanced, conservative, or liberal are also taken into consideration. Furthermore, the variabilities of error and response time, defined as the ratios of standard deviation to the mean, are computed to analyze the ability to maintain executive attention levels needed for information processing \cite{Conners2011}. 

The resting-state brain activity has been often used as the baseline for activations occurring during subsequent cognitive or sensory-motor functions. Barry \textit{et al.} suggested to consider the eyes-closed (EC) resting-state EEG as the arousal baseline for tasks not involving any visual stimuli, and the eyes-open (EO) recordings as the activation baseline for other studies including those with a visual fixation \cite{Barry2007}. Although resting-state oscillatory dynamics reflect the intrinsic activity of the brain, they have been shown to be shared by task activated networks as well \cite{Cole2014} and have been associated with performance measures during pre-stimulus and post-stimulus periods of several sensorimotor, motor learning, and attention-related tasks \cite{Reichert2015, Ozdenizci2017, Karamacoska2018, Irrmischer2017}. We thus propose that the high resolution electroencephalographic features recorded while the brain is in the wakeful and alert state can be indicators of task sustainability in a long fixed-sequence SART session. This is especially valuable since few recent studies, except for the driving simulator experiments, record data for over an hour. This analysis can be used to predict the stability of an operator's performance \textit{prior} to task execution and to adjust the interface environment, increase the number and intensity of flashes -- e.g., in the case of P300 word speller experiments, or vary the number of repetitions and type of stimulus presentation in motor imagery experiments. To analyze the readiness of the resting brain, we use multivariate pattern analysis (MVPA), a method mainly applied on the functional MRI measurements to study the connectivity of distributed brain networks involved in task-related activities \cite{Jimura2012}. This method has been previously used to identify the intrinsic and pre-trial correlates of motor learning in an EEG-based experiment \cite{Ozdenizci2017}. 

Developing regression models for such scenarios requires ground truth labels corresponding to several attention states. However, obtaining the ground truth for invisible cognitive states in general and vigilance levels in particular is a challenge \cite{BrouwerBook2015}. Several studies on vigilance assessment from simulated driving sessions label trials by visually inspecting the participants' facial features or assuming they are maximally awake and alert in the beginning of a given task and sleepy towards the end. However, our experimental results show that humans exhibit large differences in temporal transitions between their alert and sleepy states (See Figure~\ref{uni_vig}). Other protocols pause the experiment flow and ask participants to rate their own cognitive and physiological states using discrete or continuous scales \cite{Myrden2015}. These subjective evaluations are prone to high bias and experimental errors. Furthermore, momentary pauses disrupt the natural tonic levels of sustained attention in otherwise fatigued individuals \cite{Bekhtereva2014}. Finally, self-reported ratings with the Likert scale ignore the immediate reactions to the stimuli and require reflective thinking and decision making \cite{Aydin2018} while the parameters used to assess these cognitive variables should not be affected by delay and consequent memory lapses. In more objective assessments, the average number of errors gives a continuous measure for classification of vigilance patterns \cite{Gu2014}. A number of other studies on fatigue and vigilance recognition use a fusion of EEG and electrooculogram (EOG), increase in the eye closure intervals (PERCLOS), variations in the circadian rhythm, slowness of reaction time, and changes in the speech signal features, off-road gaze, and face orientation \cite{Yang2010, Craye2016, ZhangNaN2016, Zheng2017, Zheng2019}. Processing these physiological events requires extra modules and may not be realistic in all settings. Thus, an automatic method for quantifying the ground truth of vigilance levels through objective measures, such as the error rates and response time, deems essential. 

In clinical assessments of attention deficit and hyperactivity, \textit{variations} in the number of errors and response time besides the absolute values of these measurements are considered to be informative but are not widely discussed in the literature. As a first of its kind, Loo \textit{et al.} obtained the number of errors, response time, and variability of RT and task performance --without specifying how they were calculated -- from the ADHD and control groups, and reported the task variability of ADHD patients was linearly correlated with lower levels of frontal alpha, frontal beta (17 - 18 Hz), and parietal beta power (13 - 14 Hz) \cite{Loo2009}. Karamacoska \textit{et al.} designed two pre-task, two-minute resting state sessions followed by auditory SART blocks \cite{Karamacoska2018}. Their step-wise regression models showed that changes in the power of delta and alpha bands during the task with respect to the resting-state recordings could predict the task accuracy, mean response time, and standard deviation of response time with an adjusted $R^2$ below 0.22 for each model. Jeunet \textit{et al.} analyzed a large number of psychological and neurophysiological parameters from psychometric questionnaires as well as pre-trial and trial band-power ratios to predict performance in a mental imagery-based BCI \cite{Jeunet2015}. The Blankertz's SMR-predictor, an indicator of sensory motor strength with respect to the resting states, was found the most reliable feature among the analyzed predictors. Reichert \textit{et al.} used the resting-state SMR power to predict if a participant is a responder or non-responder in a task of regulating the SMR power \cite{Reichert2015}. A template-matching approach was recently used to obtain spectral correlates between the resting-state blood oxygenation levels from fMRI scans with the EEG vigilance time series, defined as the ratio of alpha to delta plus beta powers. Vigilance in this case was only assessed during the EO and EC states \cite{Falahpour2018}.

In this study, we characterize the neural correlates of objective measures for attention and fatigue based on the spatio-spectral EEG features from eyes-open and eyes-closed resting state recordings prior to a long SART session. Our contributions are multi-fold: First, a fully automated pipeline is presented for preprocessing, artifact removal, and feature extraction from the pre-task EEG signals. Second, a novel cumulative vigilance score (CVS) is calculated from the error rates and hit response time (HRT) of correct non-target trials, and is \textit{adapted} to the reaction time during the initial 50 seconds when the participants are still highly attentive. We emphasize modeling of HRT and CVS variability in addition to their average values as indicators of performance consistency or stability for sustaining attention and motor execution. Third, we train various neural networks with one fully-connected layer for linear regression from band-power ratios, and obtain the relative weights of hidden units to automate identification of the significant spatiospectral features. The relative rankings of obtained weights uncover associations between the performance measures with different beta sub-bands and wide-band gamma. Finally, we develop multivariate regression models using a thorough feature relevance analysis, and demonstrate how small feature subsets from pre-task, intrinsic networks are successful in predicting the overall task-related performance measures. These experimental setup and vigilance scores were previously used in a substantially different work that focused on predicting the performance measures from phase synchrony features obtained during the 8-minute long blocks of the actual sustained attention task \cite{Torkamani-Azar2019a}. 

The rest of this paper is organized as follows. Section~\ref{Methods} is dedicated to the explanation of our automated feature extraction and scoring pipelines and to the development of neural networks and regression models. Section~\ref{Results} presents the obtained results and significant predictors of the continuous performance measures. Finally, Section~\ref{Discuss} includes a comparison of major findings with the literature and their implications for predicting the users' abilities in sustaining their vigilance and reaction time from the pre-task EEG.

\section{Materials and Methods} \label{Methods}
\subsection{Participants}
Ten healthy and right-handed participants (six female and four male) with the average age of 30.25 $\pm$ 6.95 (min: 22, max: 45.5) attended the fixed-sequence SART sessions. In the beginning of each session, participants were asked about their sleep patterns in the last 24 hours, consumption of caffeinated products, and use of drowsiness-inducing medications. Participants generally attended the experiments in the early afternoon hours as meal consumption would induce fatigue and drowsiness in idled brain networks during a repetitive task. Except for two individuals who had attended a multi-mode SART session two years ago, all participants were na\"ive to BCI experiments in general and to the SART protocol in particular. All participants provided signed informed consents and received monetary compensation upon experiment completion. The recruitment and experimental procedures were approved by the Sabanci University Research Ethics Council in February 2016.

\begin{figure}[b]
\captionsetup[subfigure]{justification=centering}
\centering
\begin{subfigure}{0.24\textwidth}
\includegraphics[width=.9\linewidth, angle=0]{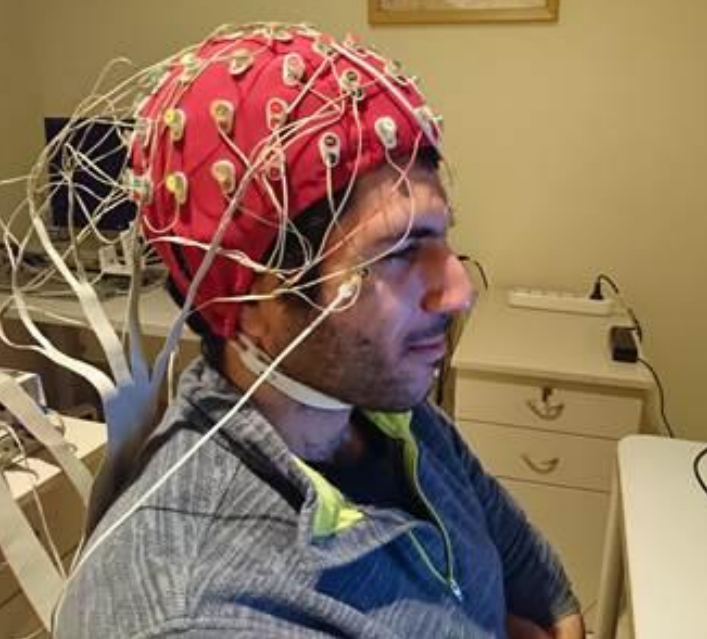}\caption{}
\end{subfigure}
\begin{subfigure}{0.24\textwidth}
\includegraphics[width=0.98\linewidth, angle=0]{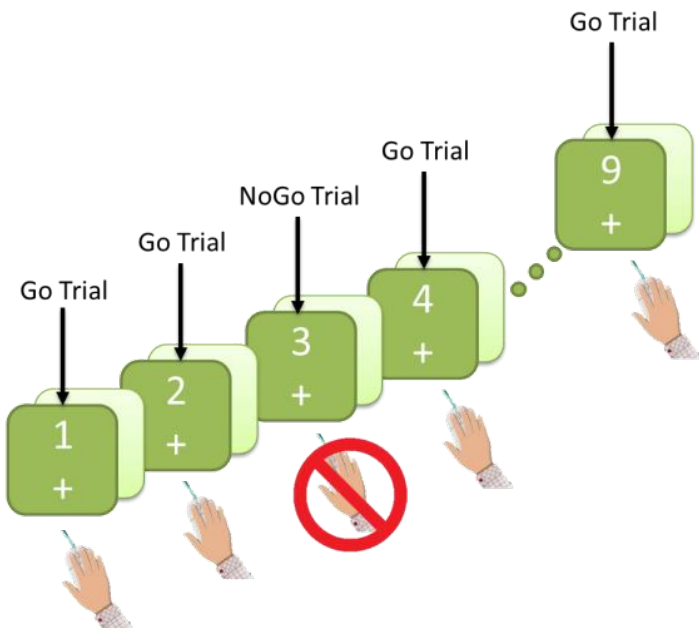} \caption{}
\end{subfigure}
\caption{(a) A 64-channel Biosemi headset (Biosemi Inc., Amsterdam, The Netherlands) and 3 surface electrodes for simultaneous EOG recording. (b) One sequence of the fixed-SART-varying-ISI paradigm. Digit display duration: 250 ms, response interval: 300 ms, fixation duration (ISI): randomly varying between 400 and 1000 ms.}\label{uni_SART_flow}
\end{figure}

\subsection{Procedure}
The task session consisted of 12 blocks of fixed-SART paradigm as described in \cite{Dockree2005}. Each block included 25 sequences of digits from 1 to 9 appearing sequentially with varying inter-stimulis intervals (ISI). Participants were asked to press the mouse left button once and as soon as they saw any digit appearing on the screen, except for the digit 3, in which case they would have to withhold their responses. Thus, digit 3 was specified as the target while the other eight digits were considered as non targets. Plot (b) of Figure~\ref{uni_SART_flow} shows one sequence of this experiment. ISIs were randomized to eliminate any chance of participants becoming habituated by the stimulus timing and to reduce the occurrence probability of automatic clicks. Their duration was also longer than the ISIs in similar SART experiments to ensure a relatively free state of mind before starting a new trial. Each block lasted between 8:04 and 8:19 minutes, and there were 5-second relaxation periods after each block. A full session of this SART paradigm would thus last for 2700 trials and around 100 – 105 minutes. A practice session with one instance of digits from 1 to 9 appearing randomly on the screen was first conducted. To have a baseline for each individual's brain activities, all participants completed a 2.5-minute resting state with eyes open (EO) followed by a 2.5-minute resting state with eyes closed (EC). We attempted to ensure participants were in alert and wakeful states prior to the task; as such, the interface would prompt them to perform a specific mental multiplication operation \textit{before} each resting-state session. 

\begin{figure*}[]
\captionsetup[subfigure]{justification=centering}
\centering
\includegraphics[width=0.85\linewidth, angle=0]{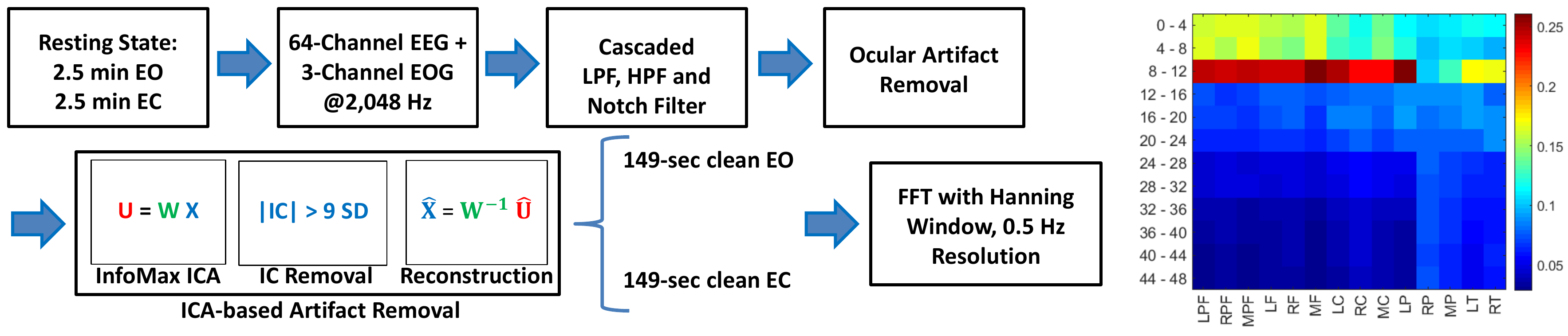}
\caption{Automated pipeline for recording, preprocessing, and feature extraction from the pre-task resting-state EO and EC EEG signals. Signals were band-pass filtered from 1 to 70 Hz and notch filtered at 50 Hz. Ocular artifacts were removed using the linear method explained in \cite{Saa2011a}. Independent components (ICs) of the logistic Infomax algorithm \cite{Bell1995} from the EEGLAB toolbox \cite{Delorme2004} were z-score standardized before artifact rejection. The heat map demonstrates ratios of band-power features from the pre-task, EO session of one participant (S10) for the following ROIs: left, midline, and right pre-frontal (LPF, MPF, and RPF), frontal (LF, MF, and RF), central (LC, MC, and RC), parietal (LP, MP, and RP), and left and right temporal (LT and RT) regions. See Section~\ref{eeg_proc} for more explanation.}\label{procedure_resting}
\end{figure*}

\subsection{EEG Analysis and Feature Extraction}\label{eeg_proc}
Data collection was performed in a dimly lit EEG room within a Faraday cage. Participants were comfortably seated in a chair 20 cm away from a 17-inch LCD monitor. Monopolar EEG activity was collected via 64 Ag/AgCl active electrodes mounted according to the 10-20 International Electrode Placement System as shown in plot (a) of Figure~\ref{uni_SART_flow}. The ActiView software was used for data recording at 2048 Hz. The fully automated preprocessing and feature extraction steps applied on the resting-state EO and EC signals were performed offline, as described in Figure~\ref{procedure_resting}. 

Defining narrow bands especially for beta oscillations used in correlation and regression models allowed us to account for individual traits in modulation of different frequencies and to better analyze the opposite roles of lower beta frequencies --as indicators of fast idleness, middle beta oscillations which appear during high engagement and alertness, and the higher beta activities which reflect signs of existing anxiety. Since our preliminary analysis had shown participants had different levels of base-band powers in the eyes-open and eyes-closed resting states and the actual task sessions, we computed the ratios of the 12 non-overlapping band powers for each session and each trial from the magnitudes of the FFT coefficients. Comparing these ratios across participants enabled us to analyze individual differences in a unified manner. To more easily study spatial variations in cortical activities, our 64 electrodes were grouped into 14 regions of interest (ROIs) as mentioned in Figure ~\ref{procedure_resting}. These 14 x 12 features are hereafter referred to as the BP-ROI feature set and a heat map of these ratios from the pre-task, EO state of participant S10 is demonstrated in the same figure.

\subsection{Performance Measures}\label{perf_measures}
\begin{figure*}[]
\captionsetup[subfigure]{justification=centering}
\centering
\includegraphics[width=0.85\linewidth, angle=0]{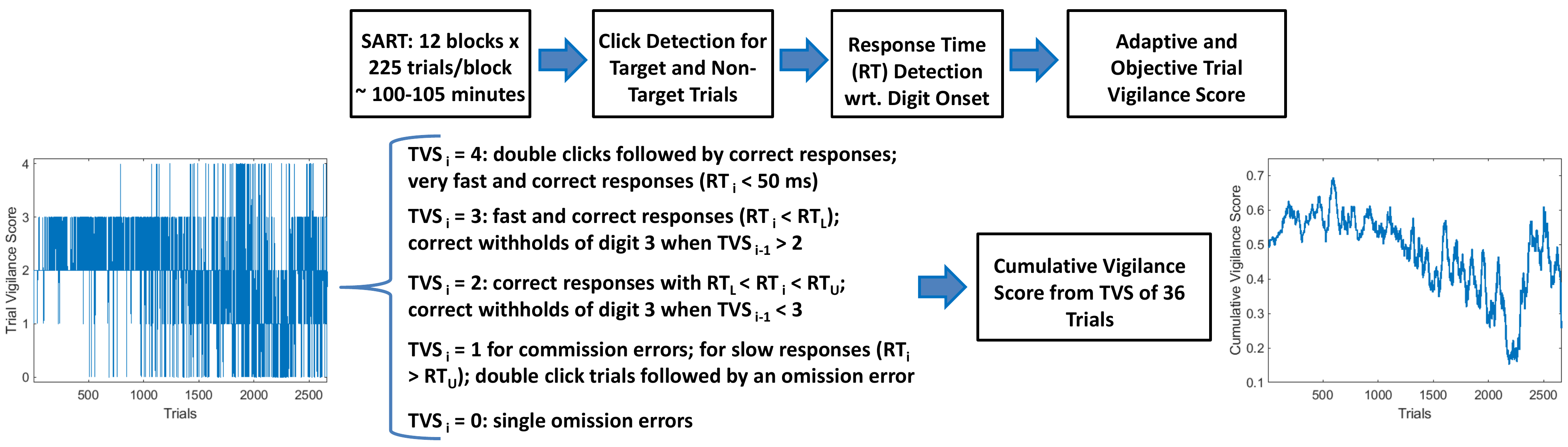}
\caption{Pipeline for detection of trial-wise events and calculating the adaptive and objective Trial Vigilance Score (TVS) and Cumulative Vigilance Score (CVS). During the experiment, trials are automatically labeled according to the digit type --target or non-target-- and click detection. In the 5-label TVS scheme, for each trial $i$, RT is compared with $RT_L$ = 250ms and $RT_U$ = mean+2 standard deviation of RTs from the first 27 trials. CVS variability during the long SART experiment indicates the performance inconsistency (inability to maintain a consistent attention and performance level). See Section~\ref{perf_measures} for more explanation.}\label{procedure_task}
\end{figure*}

Figure~\ref{procedure_task} demonstrates two different types of designed performance measures, trial-based and cumulative, and summarizes how the visual interface determines the trial response time (RT) and detects the occurrence of commission errors (CE) during target (NoGo) trials, omission errors (OE) during the non-target (Go) trials, and double clicks. CE\% and OE\% are subsequently defined as the number of CEs and OEs divided by the total number of target and non-target trials in completed blocks. Trial response time is defined as the latency of each click with respect to the digit onset. Note we do not omit trials with response time below a certain threshold as done in \cite{Karamacoska2018} since we are interested in analyzing these fast reactions as a natural occurrence in the response traits of our participants. Section~\ref{discussDelta} includes a comparison of our findings on neural correlates of impulsivity with those from studies which omitted these fast responses. Next, the hit response time (HRT) is defined as the response time for correct Go trials, i.e., trials with non-target digits for which a correct click was detected. Since variations in response time indicate the inability to maintain vigilance during long attention tasks and tests of attention deficits, variability of the overall HRT was calculated as the ratio of the standard deviation to the averaged HRT \cite{Torkamani-Azar2019a}. 

\subsubsection{Adaptive Vigilance Labels}\label{adaptive}
To come up with a continuous and objective measure for labeling sustained attention without interrupting the users, we propose an adaptive, 5-level Trial Vigilance Score (TVS) algorithm as described in Figure~\ref{procedure_task}. To avoid penalizing participants with conservative and slower responses, we adjusted the upper threshold to accommodate for each person’s response style assuming that reactions in the first 27 trials before occurrence of fatigue signs are generally faster. This definition of TVS includes correct response commission and inhibition while rewarding consistency in correct and fast performance (levels 2 to 4), and penalizing inconsistencies when double clicks are performed and subsequent trials are missed (level 1). Double or triple click events usually occur prior to omission errors when the participant misses the natural flow of trials and automatically clicks before onset of a new digit due to being in a low vigilant state. Less frequently, these events take place when a participant is in a high vigilant state and expects the next digit, but mistakenly clicks due to the varying duration of the ISI while still managing to respond correctly to the next digit. This novel labeling strategy thus provides a useful and adaptive measure for assessment of vigilance maintenance in this task.

We obtained the Cumulative Vigilance Score (CVS) at each trial by calculating the average TVS from 36 preceding trials -- lasting for 4 sequences or 73 seconds, and normalizing the result between 0 and 1. Subsequently, performance measures analyzed in this paper consist of (a) the discrete-valued task errors: CE\% and OE\%, (b) the continuous task scores: averages of the overall CVS (CVSmean) and HRT (HRTmean) which denote the score and speed of task performance, and (c) the continuous task consistency measures: variability of the overall score (CVSvar) and speed (HRTvar) as indicators of stability and sustainability -- or lack thereof -- in attending to the long SART. 

\subsection{BP-ROI Feature Selection and Visualization with Neural Networks}
The literature contains several discussions on correlations among the performance measures with channel-wise band-power features. Due to the small size of our dataset and sensitivity to the data of individual participants, we instead investigated the use of neural networks with one fully-connected layer composed of multiple hidden units for developing the aforementioned regression models. This analysis will open the path for analyzing the learned weights for feature extraction and reduction for automated vigilance estimation in the context of BCIs. Zheng \textit{et al.} had investigated the critical frequencies and electrodes from trained deep belief networks (DBNs) for emotion classification \cite{Zheng2015a}. They noticed beta and gamma features had received higher average weights in the trained networks across all participants, and saw an improvement in classification accuracy using the differential entropy of all bands with reduced electrode sets. 

Focusing on our BP-ROI features with a space dimension of 14 (ROIs) $\times$ 12 (bands), eight schemes were analyzed to predict our four continuous performance measures from each of the EO and EC states separately. Feature matrices $X_{EO}$ and $X_{EC} \in R^{168 \times N}$, $N$ being the number of participants, were separately fed to the neural networks (NN) as follows. We ran 10 dataset permutations, each consisting of leave-one-subject-out cross-validations (LOO-CV) among 10 participants for the CVSmean, HRTmean, and HRTvar measures, and 9 participants for CVSvar. S04 was removed from the CVSvar experiments since their score was more than 2 standard deviations (SD) larger than the average CVS variability. Feature matrices were standardized before being fed to the NNs that consisted of an input layer, one fully connected (FC) layer with hidden perceptrons, a rectified linear unit (ReLU) activation layer, and an output regression layer with the mean-squared-error loss function. 

Experiments were run with different number of hidden units in the FC layer. The network was initially tested with 40 hidden units with 1,000 epochs and a mini-batch size of 8, and it was noticed that although the training loss was decreasing, the validation loss reduced in the first few epochs before quickly surpassing the training error. To tackle the overfitting problem, a validation patience scheme was utilized to stop the training if the validation loss did not improve after one epoch. Using Adam \cite{Kingma2014}, a method for adaptive moment estimation as the optimization algorithm, a grid search was performed to optimize each loss function for 15 learning rates and 15 $\ell_2$ regularization coefficients that were logarithmically increased within the [$10^{-5}, 10^{-1}$] and [0.01,10] intervals, respectively. For each combination of the learning rate and $\ell_2$ parameters, the network performance was assessed in each fold with the root-mean-squared error (RMSE) between the true and predicted outputs. The LOO-CV estimator was then obtained by minimizing the average error across all the folds and permutations \cite{Lemm2011}, that is to say, 

\vspace{-0.3cm}
\begin{equation}
err_{LOO-CV}(lr, \lambda) = \dfrac{1}{MN} \, \sum_{m = 1}^{M} \, \sum_{n=1}^{N} (y_n - f(X_n|D \backslash D_n))^2
\end{equation}

\noindent Here, $M$ is the number of permutations, $N$ is the number of validation folds, $D$ and $D_n$ denote the original sample set and the validation set from the $n$-th run, $y_n$ and $X_n \in R^{168 \times 1}$ represent the true label and feature vector for sample $n$, $f(X_n|D \backslash D_n)$ is the estimated output by the neural network, and $lr$ and $\lambda$ are the learning rate and the $\ell_2$ regularization coefficients, respectively. 

To study which features were given higher priority during the training and, subsequently, to perform supervised feature selection, the input weights of the first FC layer obtained during validation from the optimal pair of our hyperparameters were summed over all the $U$ hidden units and averaged for all the $N$ folds and $M$ permutations. In other words,

\vspace{-0.3cm}
\begin{equation}
\bar{W_j} (lr^*, \lambda^*) = \dfrac{1}{M} \, . \dfrac{1}{N} \, \sum_{m = 1}^{M} \, \sum_{n=1}^{N} \, \sum_{i=1}^{U} W_{ij} (lr^*, \lambda^*),
\end{equation}

\noindent where $(lr^*, \lambda^*)$ represents the hyperparameters that minimized $err_{LOO-CV}$, and $W_{ij}$ is the weight associated with unit $i$ on the first hidden layer and feature $j$ of the input vector. The averaged weight vector $\bar{W_j} \in R^{168 \times 1}$ is subsequently visualized for feature selection. Results of these analyses are reported in Section~\ref{NN_results}.

\subsection{Resting-state Feature Relevance Analysis for Multivariate Prediction of SART Performance Measures}\label{pred_method}
Next, eight prediction schemes were designed to predict the four continuous dependent variables from the BP-ROI of the EO and EC states. Single linear regression (SLR) models were developed for the initial feature selection: each of the 168 BP-ROI features were standardized and individually entered in the model as an independent variable. After performing a LOO-CV scheme across all participants, the $R^2$, adjusted $R^2$, RMSE, Pearson's linear correlation coefficient ($r$), and its corresponding $p$-value were calculated from the true and predicted outputs of these SLR models.

Upon detecting the $n$ features whose individual regression models had resulted in prediction correlations significant at the 0.1 level for each performance measure and feature set, the possibility of predicting the studied performance measure from a group of these significant features was investigated. To identify the most predictive feature subset, a multivariate pattern analysis (MVPA) approach was utilized, and multivariate linear regression (MLR) models were trained from all the $2^n$-1 non-empty subsets \cite{Ozdenizci2017}. To test the significance of each model, a null hypothesis of having no correlation between the true and predicted outputs was defined, and features were permuted 500 times across all participants. This resulted in obtaining the $p$-value for each prediction model having achieved a correlation higher than the original feature assignment to the performance measure under investigation. Finally, for each subset, the same goodness-of-fit metrics were calculated, and models with the most outstanding metrics were reported for each of the eight prediction schemes in Section~\ref{pred_results}.

\section{Results} \label{Results}

\subsection{Behavioral Results}\label{uni_behavior_results}
Participants had an average CE\% of 17.92$\pm$8.21 and an average OE\% of 8.99$\pm$10.46. The average response time to non-target digits, HRTmean, was equal to 430.95$\pm$95.86 ms and the variability of HRT was equal to 0.53$\pm$0.20 in average. A higher HRT variability is an indicator of large differences in the trial-wise reaction time and higher variations in sustained attention. The CVSmean reflects both the performance errors and response time and was equal to 0.43$\pm$0.07 in average. Finally, an average CVSvar of 0.13$\pm$0.10 was obtained.

\begin{figure}[h]
\captionsetup[subfigure]{justification=centering}
\centering
\begin{subfigure}{0.24\textwidth}
\includegraphics[width=0.85\linewidth, angle=0]{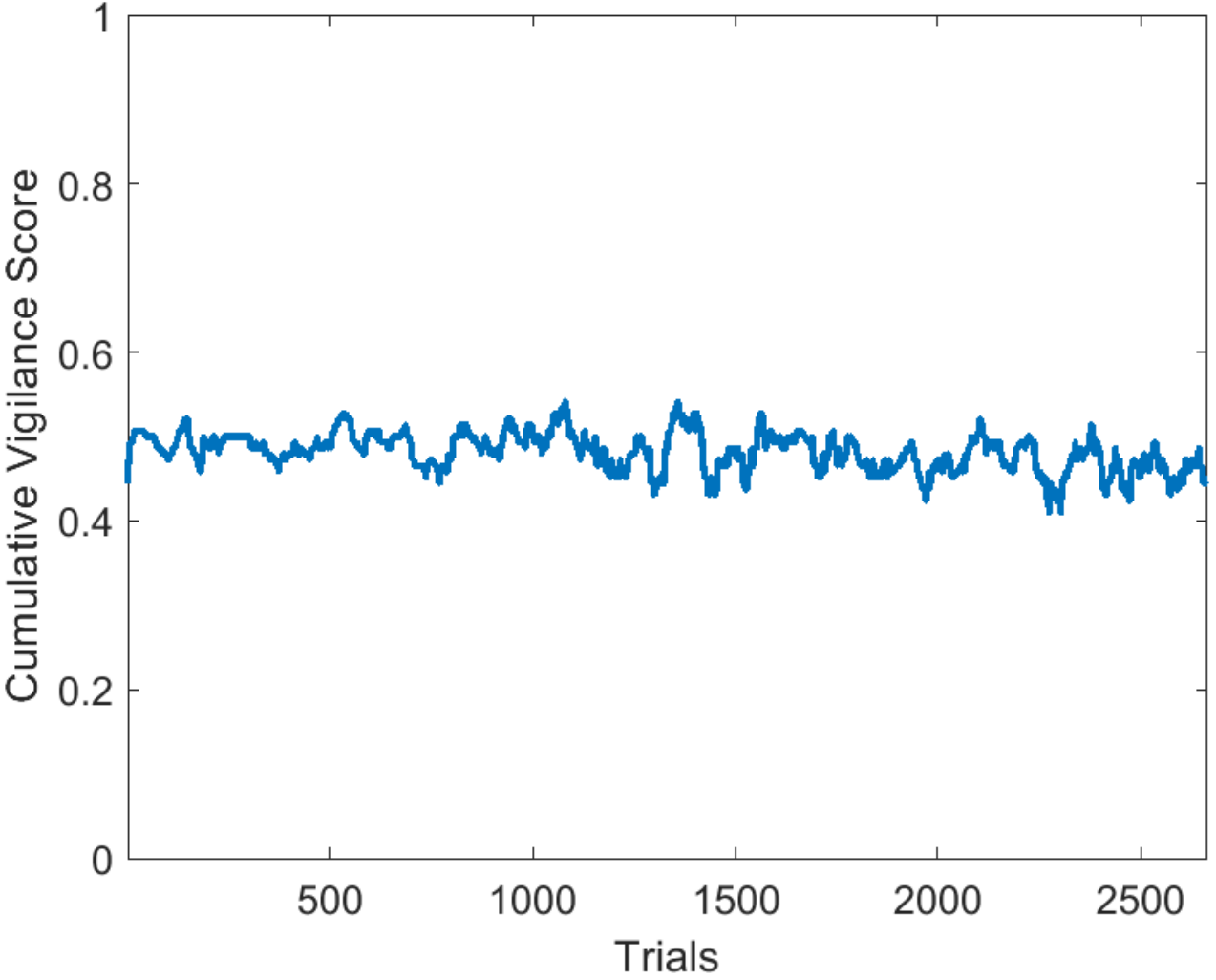}\caption{S03}
\end{subfigure}
\begin{subfigure}{0.24\textwidth}
\includegraphics[width=0.85\linewidth, angle=0]{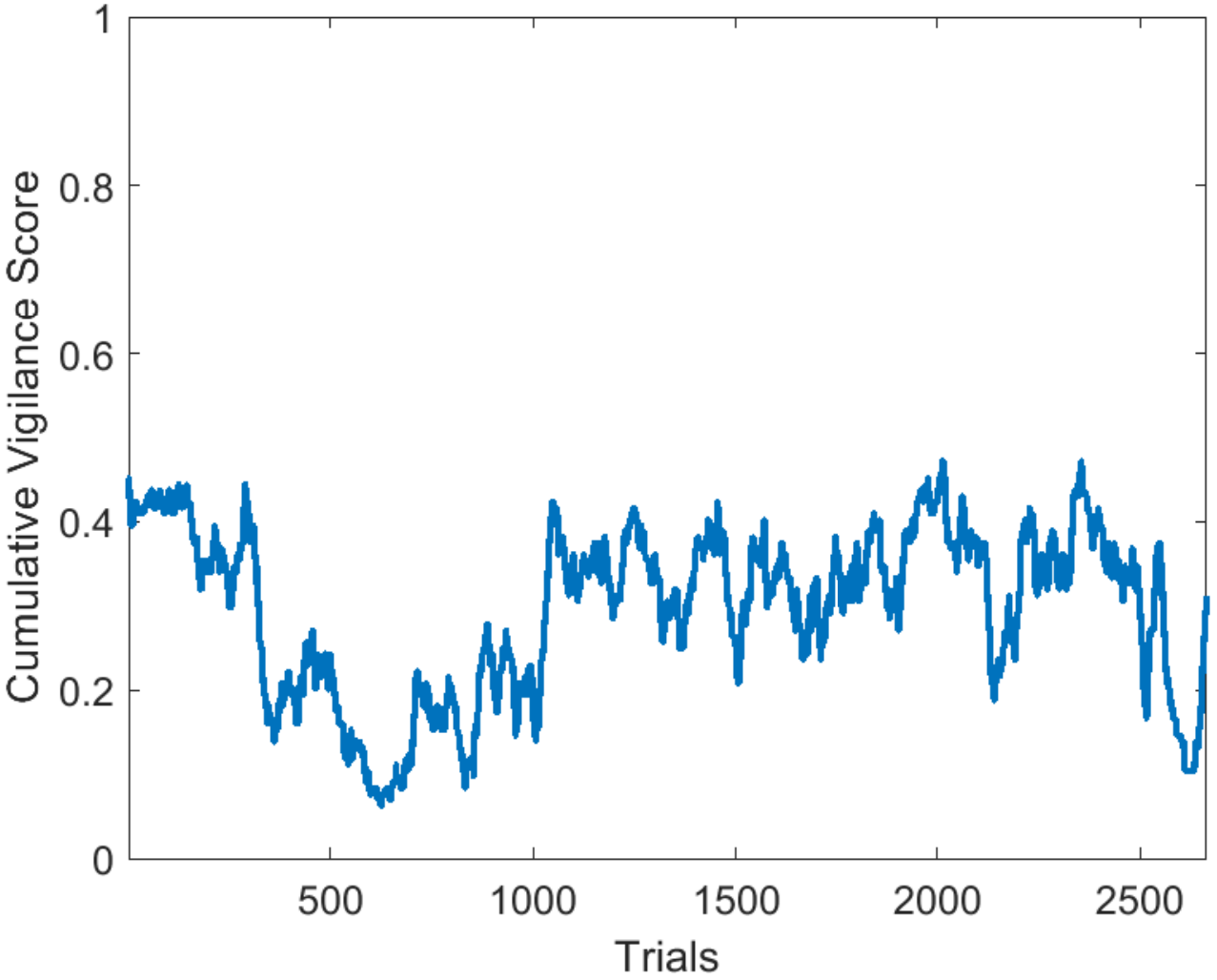} \caption{S04}
\end{subfigure}
\\
\begin{subfigure}{0.24\textwidth}
\includegraphics[width=0.85\linewidth, angle=0]{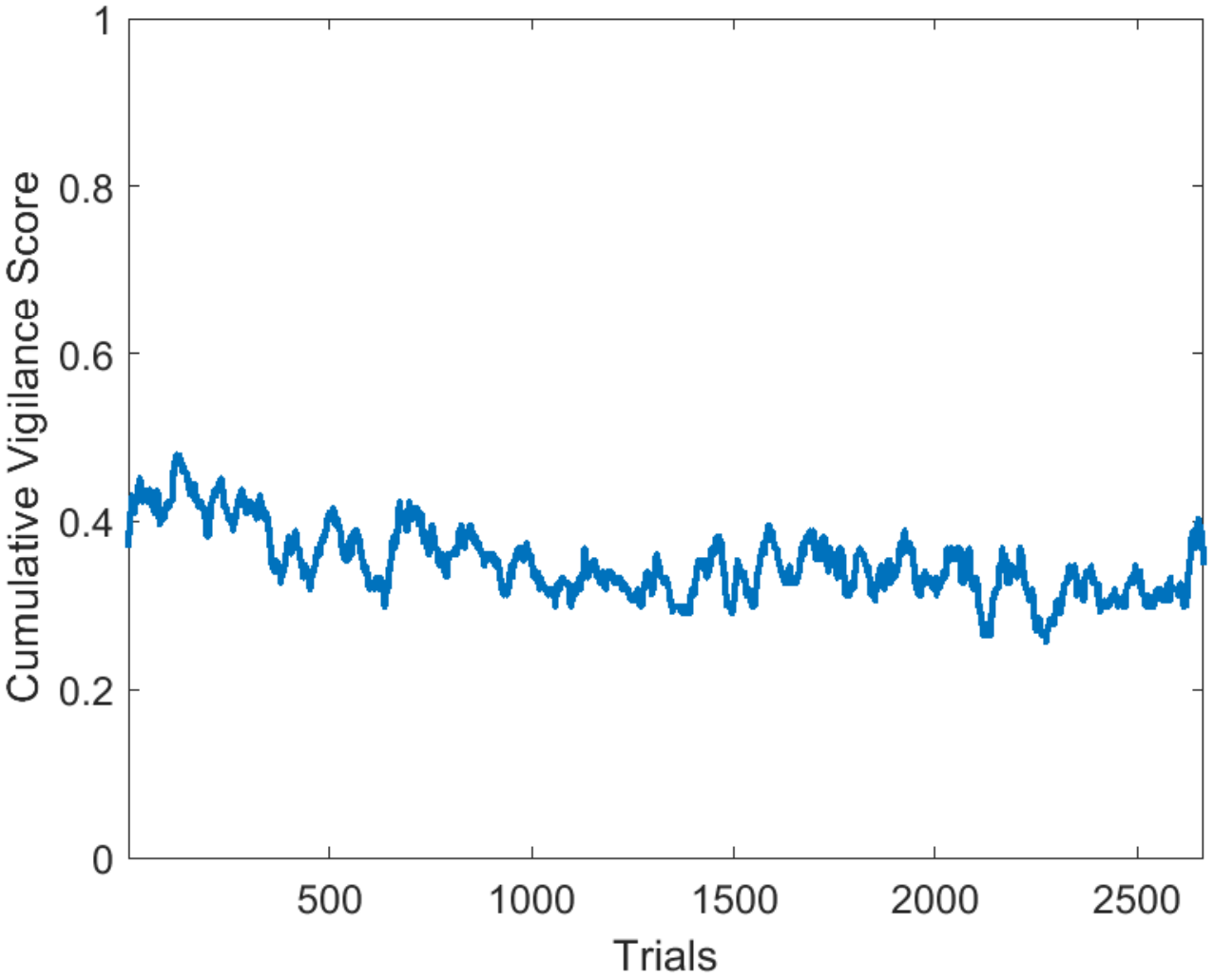} \caption{S06}
\end{subfigure}
\begin{subfigure}{0.24\textwidth}
\includegraphics[width=0.85\linewidth, angle=0]{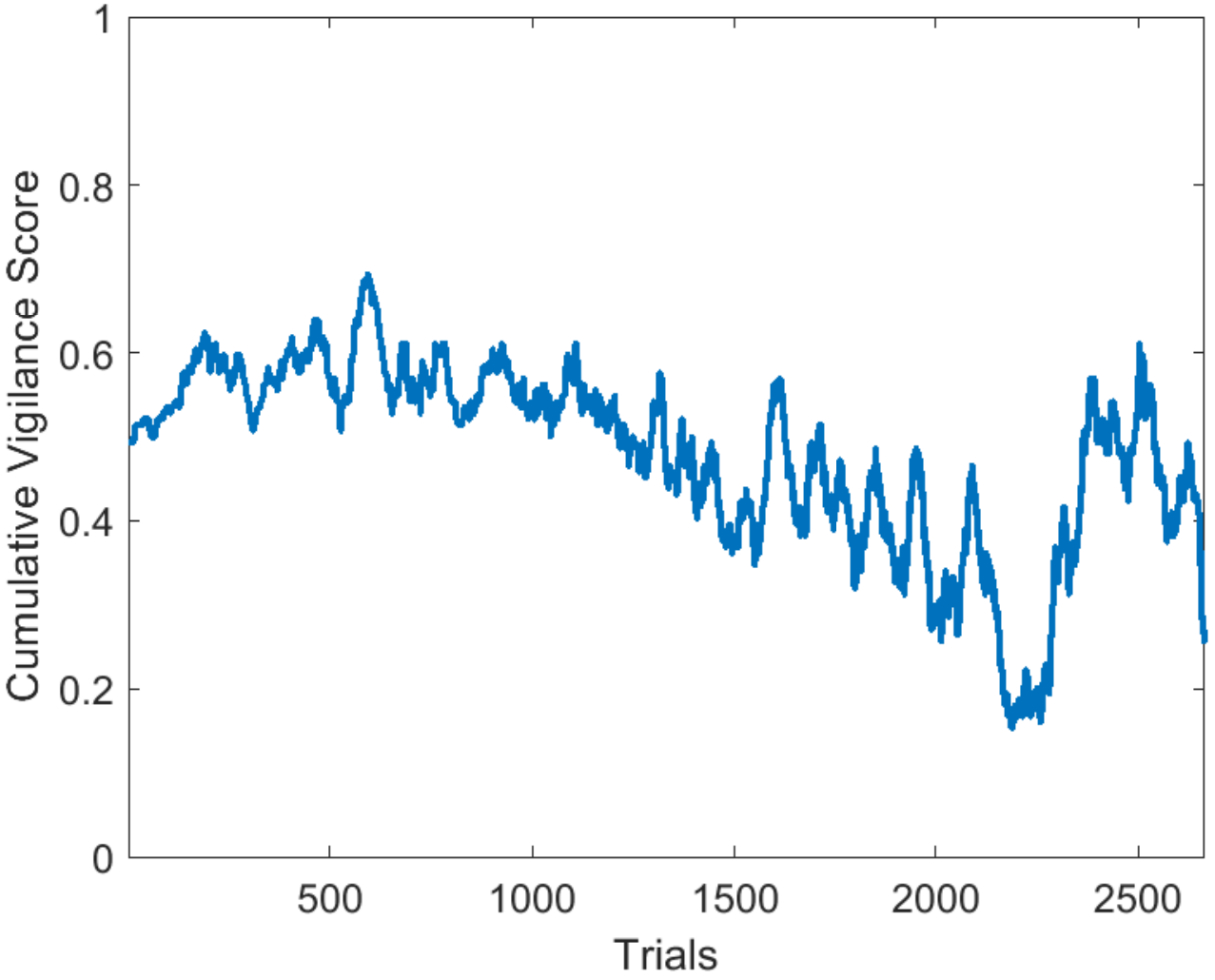} \caption{S10}
\end{subfigure}
\caption{Large inter-individual variability in terms of sustaining the performance and tonic attention, here visualized using the adaptive cumulative vigilance score (CVS), during our 105-minute SART session. More explanation in Section~\ref{uni_behavior_results}.} \label{uni_vig}
\end{figure}


To visualize these inter-individual differences in attention maintenance using a single a single time-series, Figure~\ref{uni_vig} shows the CVS curves for four sample participants: S03 with their balanced response style and consistent performance (CVsmean = 0.48, CVSvar = 0.05), S04 who fell asleep very early in the experiment and recovered later (CE\% = 31\%, OE\% = 31.83\%, CEmean = 0.30, CVSvar = 0.33), S06 with their slow and conservative responses and a gradual attention decline in the second half of the experiment (HRTmean = 583 ms), and S10 with an excellent performance in the beginning but the highest HRTvar (0.90) due to the extreme fatigue and drowsiness in blocks 9 and 10 before slightly recovering. Thus, contrary to a number of long experiments on vigilance estimation that divide the experiment intervals into three periods of high, middle, and low vigilance, these plots demonstrate that vigilance levels can drop at any moment during the experiment and be followed by a relative recovery. In fact, the majority of participants reported their alertness increased after a short, involuntary nap, indicating the brain's ability to regain its alertness after a period of idling and drowsiness.

\begin{table}[t]
\caption{Correlations among the overall behavioral measures of the fixed-sequence SART. *: $p<0.05$.}
\vspace{0.01cm}
\label{corr_table}
\centering
\renewcommand{\arraystretch}{1.3}
\footnotesize
\begin{tabular}{llllll}
\hline 
 & CE\% & OE\% & CVSmean & CVSvar & HRTmean \\ \hline 
OE\%	& 0.80*	& 	& 	& 	& 	 \\
CVSmean& -0.47	& -0.68*	& 	&	 &	\\
CVSvar& 0.90*	& 0.91*	& -0.69*	&	&  \\
HRTmean& 0.38	& 0.34	& -0.88*	& 0.51	&	\\
HRTvar& 0.80*	& 0.46	& -0.17	& 0.61	& 0.24 \\ \hline
\end{tabular}
\end{table}


Table~\ref{corr_table} demonstrates correlations among the objective behavioral measures. Since the current data set consisted of 10 participants, a minimum absolute value of 0.632 was needed to achieve the significance level of 0.05 for two-tail correlations. The obtained $p$-values are corrected at the 0.05 level using the False Discovery Rate (FDR) method \cite{Benjamini1995}. The number of commission and omission errors had a strong, positive linear association at 0.80. Furthermore, the average of overall CVS had a significant correlation with the number of missed trials while its variability, an indicator of performance inconsistency, was equally correlated with the percentage of both errors. While the average response time did not demonstrate any significant association with the number of errors, meaning that \textit{fast or slow responses did not necessarily imply wrong responses}, it did have a strong and negative correlation with the average CVS. The variability of response time was strongly associated with the number of commission errors while the variability of CVS and HRT fell short of being significantly correlated at just +0.61. Therefore, \textit{it seems informative to analyze the variabilities of CVS and response time as two time-series data in more depth due to their large correlations with the number of errors}.

\subsection{Resting-state Spatiospectral Feature Detection using Neural Networks}\label{NN_results}
\begin{figure*}[h]
\captionsetup[subfigure]{justification=centering}
\begin{subfigure}{0.24\textwidth}
\includegraphics[width=.9\linewidth, angle=0]{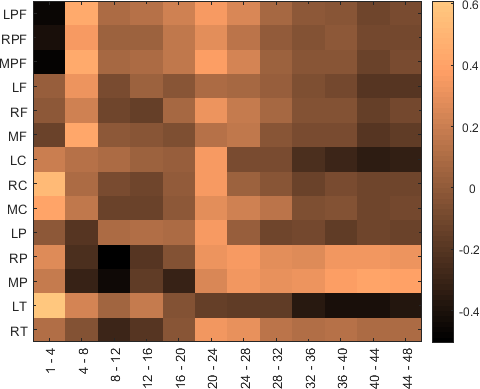}\caption{CVSmean, EO, 110 units}
\end{subfigure}
\hspace*{\fill}
\begin{subfigure}{0.24\textwidth}
\includegraphics[width=.9\linewidth, angle=0]{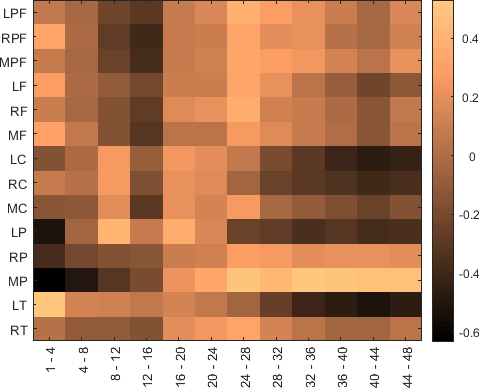}\caption{CVSmean, EC, 130 units}
\end{subfigure}
\hspace*{\fill}
\begin{subfigure}{0.24\textwidth}
\includegraphics[width=.93\linewidth, angle=0]{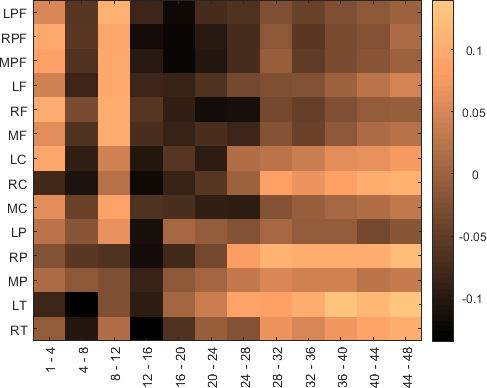}\caption{CVSvar, EO, 90 units}
\end{subfigure}
\hspace*{\fill}
\begin{subfigure}{0.24\textwidth}
\includegraphics[width=.93\linewidth, angle=0]{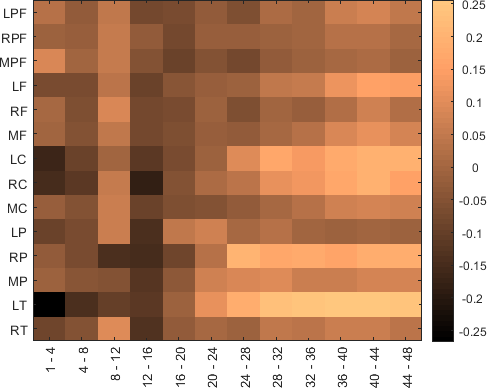}\caption{CVSvar, EC, 110 units}
\end{subfigure}
\par\medskip
\begin{subfigure}{0.24\textwidth}
\includegraphics[width=1\linewidth, angle=0]{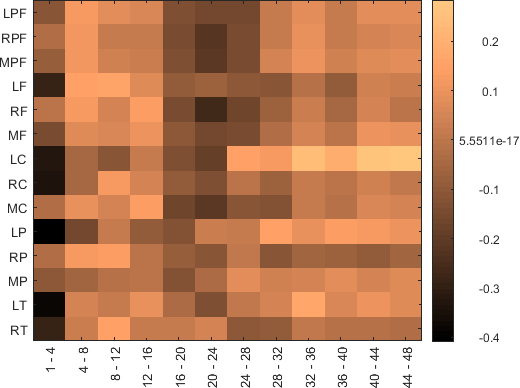}\caption{HRTmean, EO, 40 units}
\end{subfigure}
\hspace*{\fill}
\begin{subfigure}{0.24\textwidth}
\includegraphics[width=.93\linewidth, angle=0]{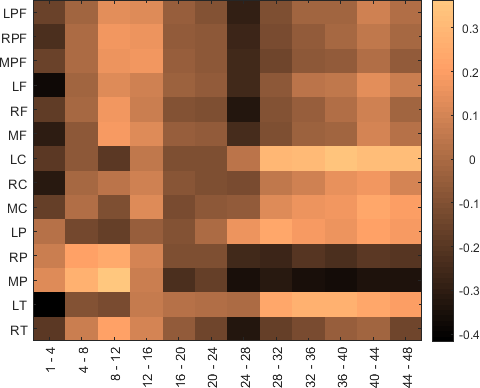}\caption{HRTmean, EC, 40 units}
\end{subfigure}
\hspace*{\fill}
\begin{subfigure}{0.24\textwidth}
\includegraphics[width=.93\linewidth, angle=0]{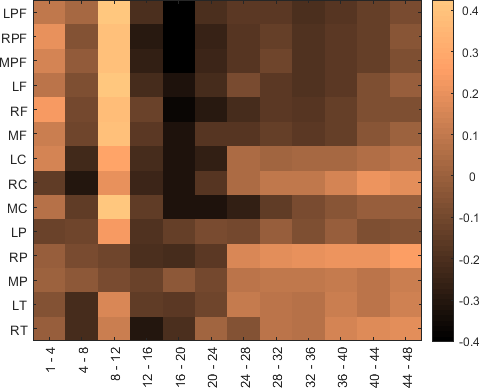}\caption{HRTvar, EO, 110 units}
\end{subfigure}
\hspace*{\fill}
\begin{subfigure}{0.24\textwidth}
\includegraphics[width=.93\linewidth, angle=0]{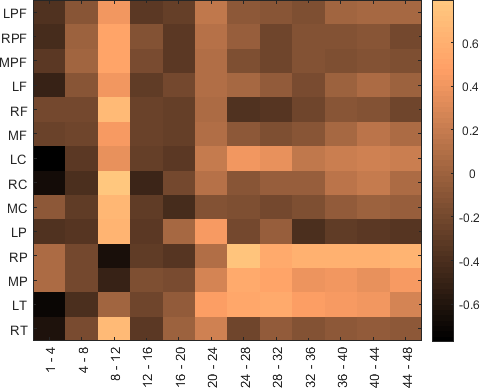}\caption{HRTvar, EC, 110 units}
\end{subfigure}
\caption{The 168-dimensional weight vectors averaged across 10 runs of one-fully-connected layer neural networks with various number of hidden units resulting in the minimum cross-validation error. Captions demonstrate the resting state and number of hidden units. L: left; R: right; M: midline; PF: pre-frontal; F: frontal; C: central; P: parietal; T: temporal. Further explanation in Section~\ref{NN_results}.}\label{uni_BP_MLPheatmaps}
\end{figure*}

One-layer neural networks were trained with normalized BP-ROI features formatted as 168-dimensional vectors for prediction of the four overall performance measures. Networks were trained for 10 runs with the LOO-CV scheme -- 9 folds for the CVSvar and 10 folds for other measures. After obtaining the pair of best regularization coefficient and learning rate for each network, $W_{ij}$, the weight associated with unit $i$ and feature $j$, was averaged over all the units to obtain $\bar{W_j}$. Figure~\ref{uni_BP_MLPheatmaps} demonstrates the heat maps for these weights averaged for all the folds and runs. The light and dark cells, respectively, denote positive and negative signs of these spatiospectral features in the cross-validated, multivariate linear regression models.

The obtained weights demonstrate the role of pre-frontal delta in predicting a lower CVS mean -- due to higher CEs and OEs-- and more inconsistent CVS during EO while increase in the left temporal delta has the opposite effect. However, they also show that the pre-frontal delta predicts faster responses and more variability in response time during EC -- a sign of hyperactivity. Frontal and central alpha from EO recordings are also confirmed to be correlates of more variability in both scores and response time. Furthermore, pre-frontal and frontal theta from EO recordings are shown to be predicting higher CVS as well as more consistent CVS and slower responses.

The heat maps also show that lower beta-1 oscillations (12-16 Hz) are generally similar to alpha in predicting slower responses and lower average CVS scores (see plots a, b, e, and f). But for predicting lower variability (more consistency) in HRT and CVS, they behave similar to the 16-24 Hz oscillations from the frontal region (plots c, d, f, and g). 
Interestingly, higher gamma ratios are not always predictors of faster or more consistent scores: During the EO and EC recordings, larger ratios of gamma from right and especially midline parietal regions are predictors of faster responses and higher CVSmean, and during the EC, predictors of shorter response time in average. The left temporal (upper) gamma, on the other hand, is more similar to the central gamma in predicting lower scores and less performance consistency from both EO and EC states, and slower responses and more variability of response time from EO features.

\begin{table*}[]
\caption{Results of the LOO-CV-based feature relevance analysis for multiple linear regression to predict the mean and variability of CVS and HRT from the EO and EC BP-ROI features. From the $n$ initially selected features for each performance measure and each feature set, all the $2^n$-$1$ non-empty subsets were individually analyzed. Statistical measures are reported for the best models of subset sizes with the highest adjusted $R^2$, highest correlation coefficient, or lowest RMSE. If more than one subset satisfied these conditions, all of the best subsets are displayed. ***: $p<0.001$, **: $p<0.01$, *: $p<0.05$.}
\vspace{0.05cm}
\label{BP_table}
\scriptsize
\centering
\begin{tabular}{c c c c c c c c | c c c c c c }
\hline
\rule{0pt}{3ex} Measure & & \multicolumn{6}{c}{BP-ROI Features, EO} &\multicolumn{6}{c}{BP-ROI Features, EC} \\ \cline{3-8} \cline{9-14}
\rule{0pt}{3ex} & & \# of features & \# of subsets & $R^2$ & Adj. $R^2$ & RMSE & Corr. rho & \# of features & \# of subsets & $R^2$ & Adj. $R^2$ & RMSE & Corr. rho\\ \hline
\rule{0pt}{3ex} CVSmean& & 3 & 20 & 0.911 & 0.867 & 0.021 & 0.956*** & 2 & 10 & 0.830 & 0.782 & 0.029 & 0.920*** \\ \hline
\rule{0pt}{3ex} CVSvar& & 1 & 4 & 0.347 & 0.254 & 0.055 & 0.618** & &  & & & & \\
& & 2 & 6 & 0.421 & 0.229 & 0.052 & 0.691** & 2 & 21 & 0.680 & 0.574 & 0.038 & 0.828***\\ \hline
\rule{0pt}{3ex} HRTmean& & 8 & 495 & 1.000 & 1.000 & 0.392 ms & 1.000*** & 3 & 20 & 0.828 & 0.741 & 37.766 ms & 0.915***\\ \hline
\rule{0pt}{3ex} HRTvar& & 2 & 28 & 0.361 & 0.179 & 0.149 & 0.640* & 2 & 3 & 0.141 & -0.105 & 0.173 & 0.486*\\ \hline
\end{tabular}
\end{table*}

\begin{figure*}[]
\captionsetup[subfigure]{justification=centering}
\tiny
\begin{subfigure}{0.28\textwidth}
\includegraphics[width=0.62\linewidth, angle=0]{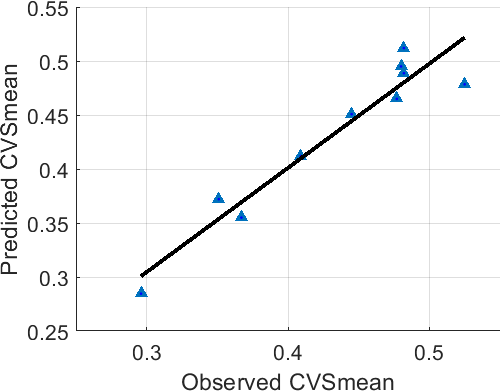}\caption{CVSmean, EO, $r=$ 0.956***.}
\end{subfigure}
\begin{subfigure}{0.20\textwidth}
\includegraphics[width=0.6\linewidth, angle=0]{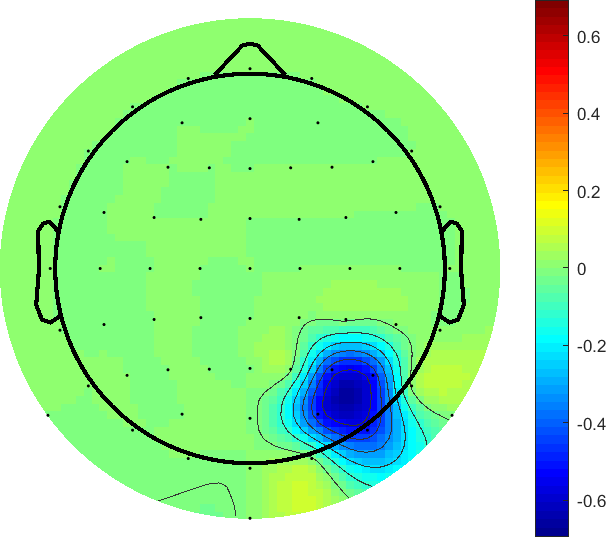}\caption{Alpha}
\end{subfigure}
\begin{subfigure}{0.20\textwidth}
\includegraphics[width=0.6\linewidth, angle=0]{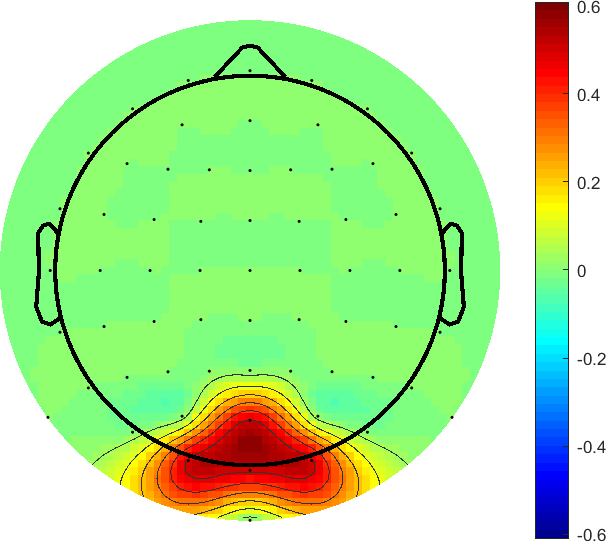}\caption{Mid-Beta}
\end{subfigure}
\begin{subfigure}{0.20\textwidth}
\includegraphics[width=0.6\linewidth, angle=0]{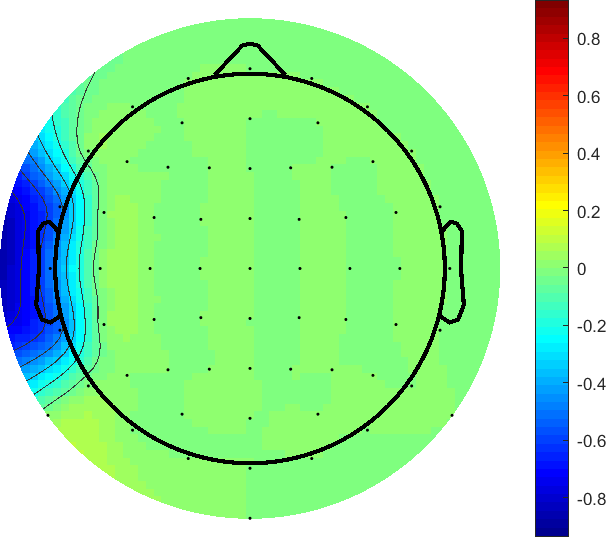}\caption{Upper Gamma}
\end{subfigure}\\
\par\smallskip
\begin{subfigure}{0.28\textwidth}
\includegraphics[width=0.62\linewidth, angle=0]{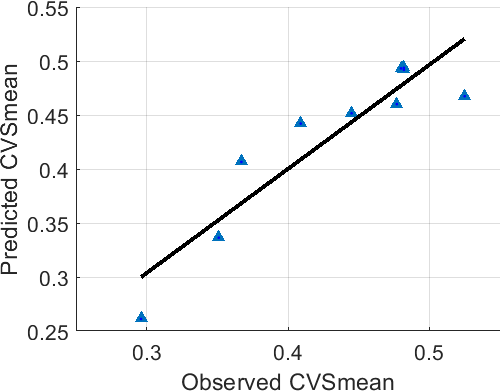}\caption{CVSmean, EC, $r=$ 0.920***.}
\end{subfigure}
\begin{subfigure}{0.20\textwidth}
\includegraphics[width=0.6\linewidth, angle=0]{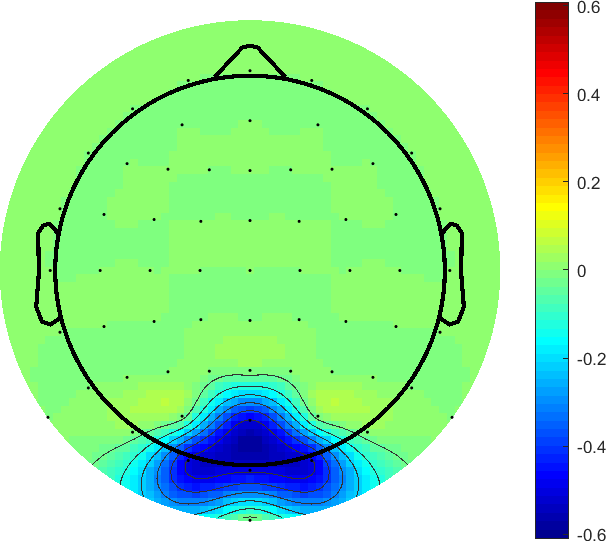}\caption{Theta}
\end{subfigure}
\begin{subfigure}{0.20\textwidth}
\includegraphics[width=0.6\linewidth, angle=0]{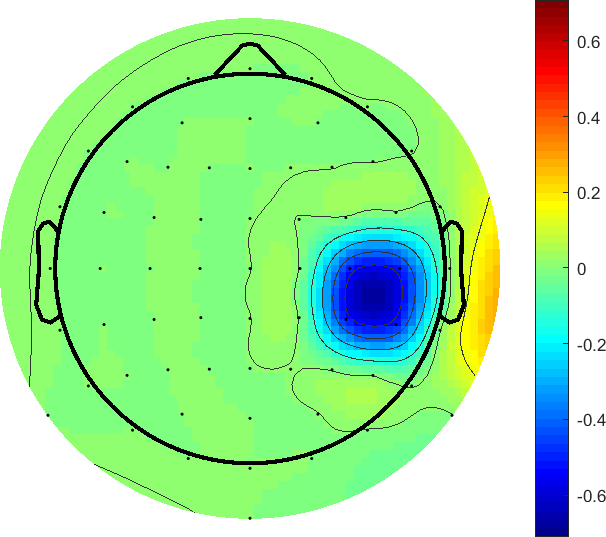}\caption{Lower Gamma}
\end{subfigure}\\
\par\smallskip
\begin{subfigure}{0.28\textwidth}
\includegraphics[width=0.62\linewidth, angle=0]{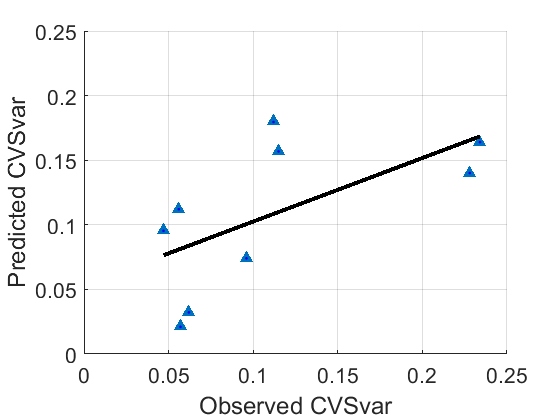}\caption{CVSvar, EO, $r=$ 0.618**.}
\end{subfigure}
\begin{subfigure}{0.20\textwidth}
\includegraphics[width=0.6\linewidth, angle=0]{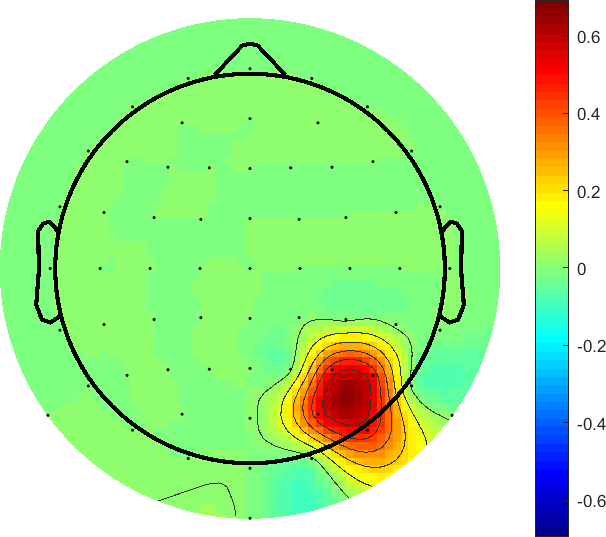}\caption{Upper Beta}
\end{subfigure}\\
\par\smallskip
\begin{subfigure}{0.28\textwidth}
\includegraphics[width=0.62\linewidth, angle=0]{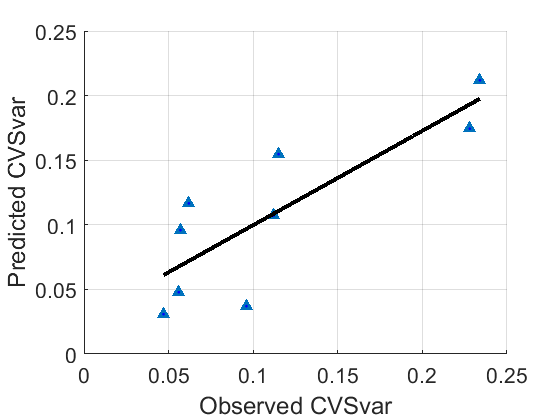}\caption{CVSvar, EC, $r=$ 0.828***.}, $p<$0.001
\end{subfigure}
\begin{subfigure}{0.20\textwidth}
\includegraphics[width=0.6\linewidth, angle=0]{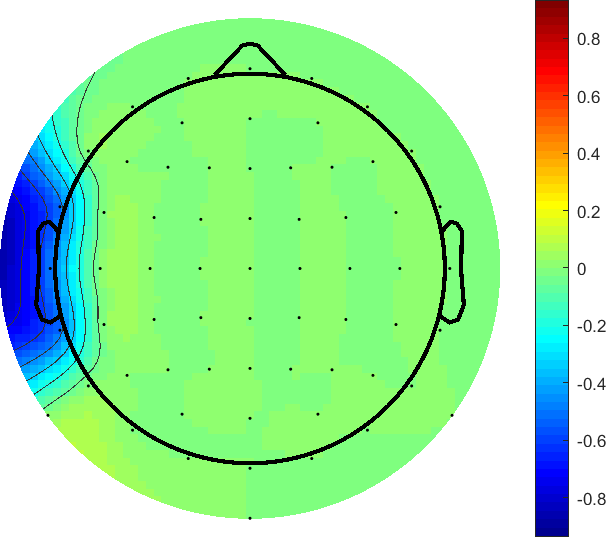}\caption{Delta}
\end{subfigure}
\begin{subfigure}{0.20\textwidth}
\includegraphics[width=0.6\linewidth, angle=0]{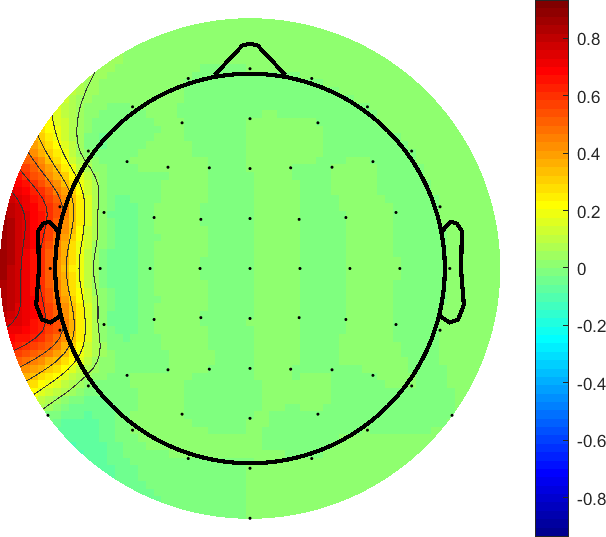}\caption{Upper Beta}
\end{subfigure}\\
\par\smallskip
\begin{subfigure}{0.28\textwidth}
\includegraphics[width=0.62\linewidth, angle=0]{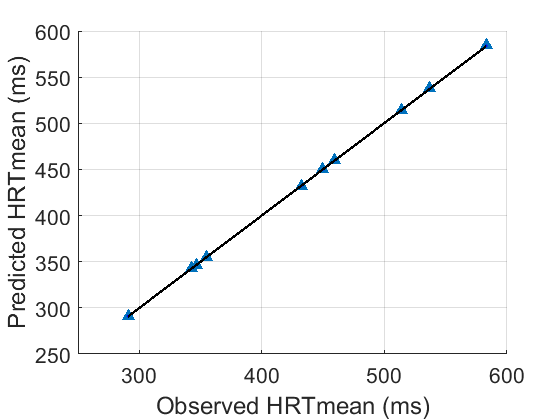}\caption{HRTmean, EO, $r=$ 1.000***.}
\end{subfigure}
\begin{subfigure}{0.20\textwidth}
\includegraphics[width=0.6\linewidth, angle=0]{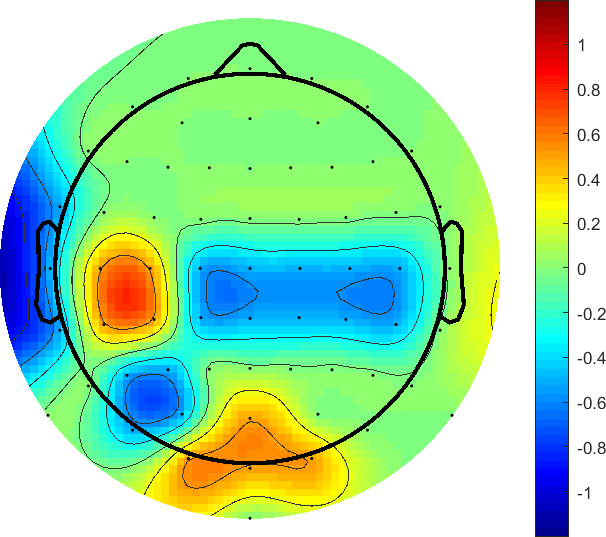}\caption{Delta}
\end{subfigure}
\begin{subfigure}{0.20\textwidth}
\includegraphics[width=0.6\linewidth, angle=0]{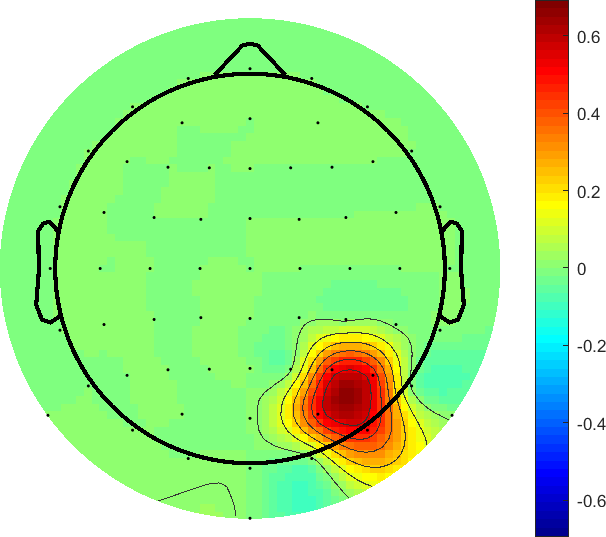}\caption{Alpha}
\end{subfigure}
\begin{subfigure}{0.20\textwidth}
\includegraphics[width=0.6\linewidth, angle=0]{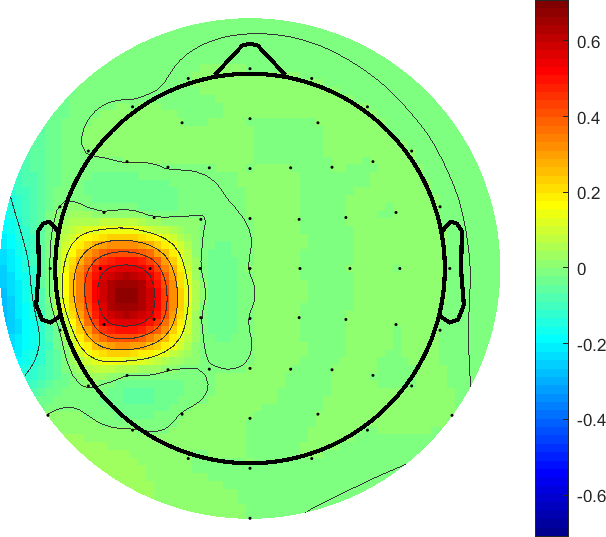}\caption{Upper Gamma}
\end{subfigure}\\
\par\smallskip
\begin{subfigure}{0.28\textwidth}
\includegraphics[width=0.62\linewidth, angle=0]{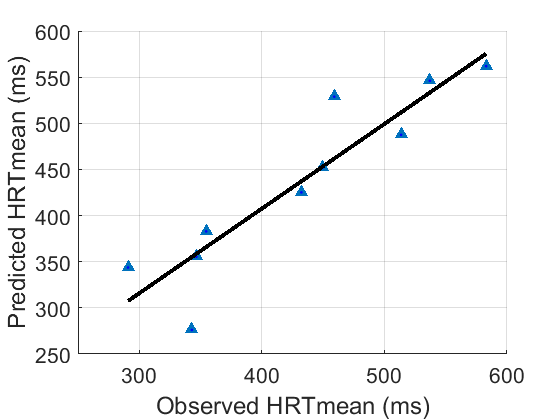}\caption{HRTmean, EC, $r=$ 0.915***.}
\end{subfigure}
\begin{subfigure}{0.20\textwidth}
\includegraphics[width=0.6\linewidth, angle=0]{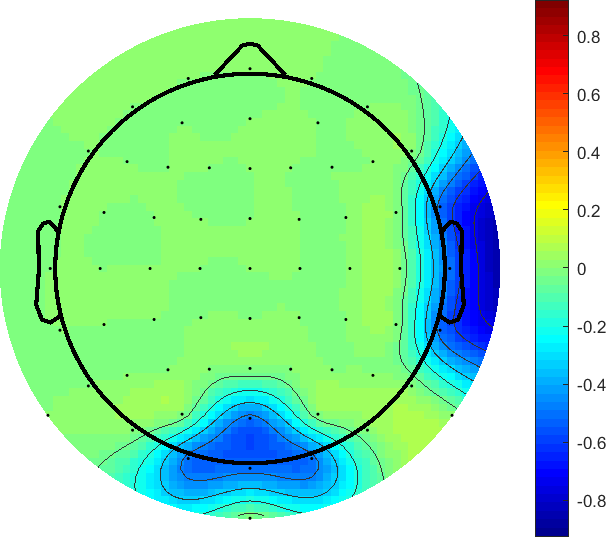}\caption{Upper Beta}
\end{subfigure}
\begin{subfigure}{0.20\textwidth}
\includegraphics[width=0.6\linewidth, angle=0]{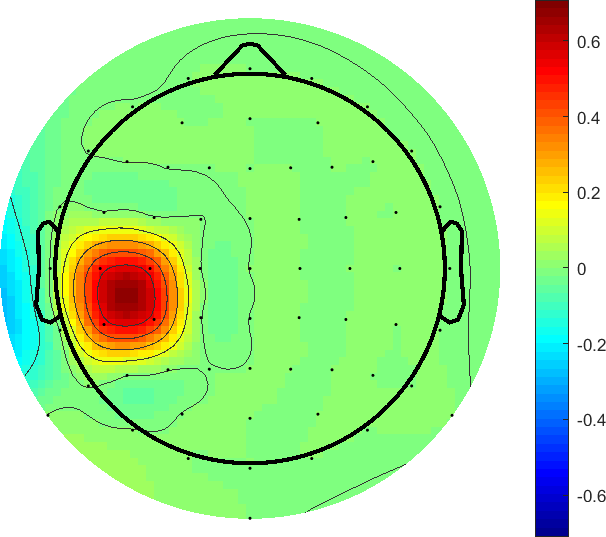}\caption{Upper Gamma}
\end{subfigure}\\
\par\smallskip
\begin{subfigure}{0.28\textwidth}
\includegraphics[width=0.62\linewidth, angle=0]{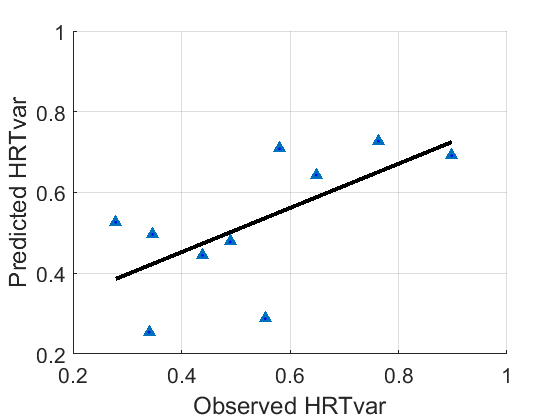}\caption{HRTvar, EO, $r=$ 0.640*.}
\end{subfigure}
\begin{subfigure}{0.20\textwidth}
\includegraphics[width=0.6\linewidth, angle=0]{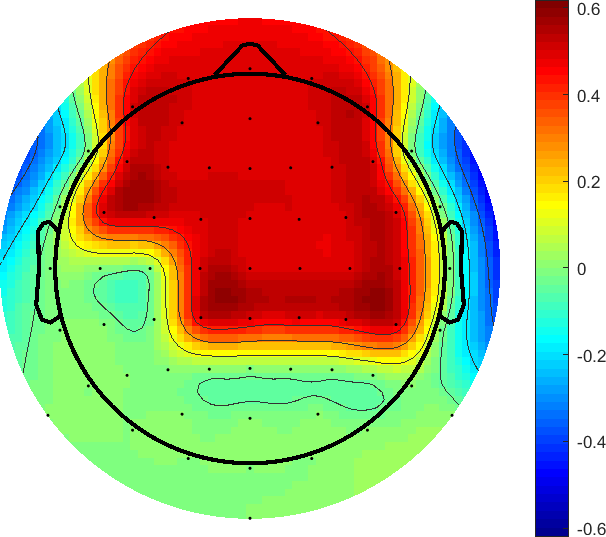}\caption{Alpha}
\end{subfigure}\\
\par\smallskip
\begin{subfigure}{0.28\textwidth}
\includegraphics[width=0.62\linewidth, angle=0]{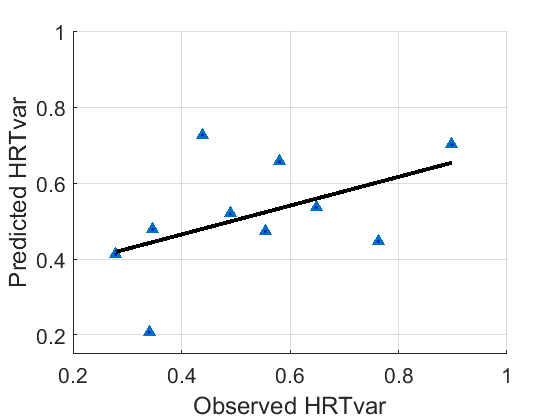}\caption{HRTvar, EC, $r=$ 0.486*.}
\end{subfigure}
\begin{subfigure}{0.20\textwidth}
\includegraphics[width=0.6\linewidth, angle=0]{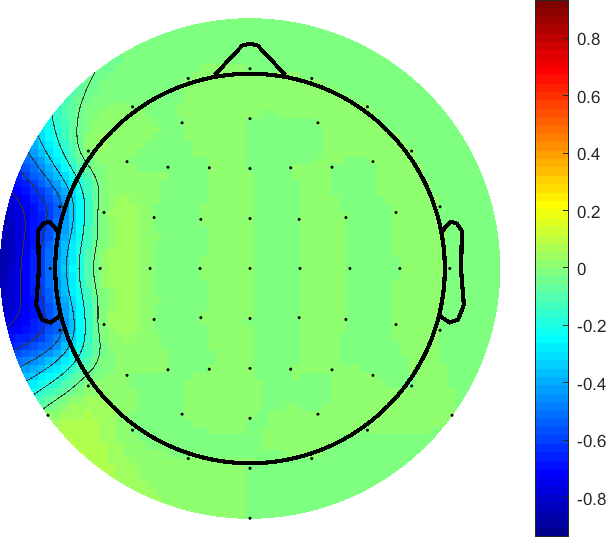}\caption{Delta}
\end{subfigure}
\begin{subfigure}{0.20\textwidth}
\includegraphics[width=0.6\linewidth, angle=0]{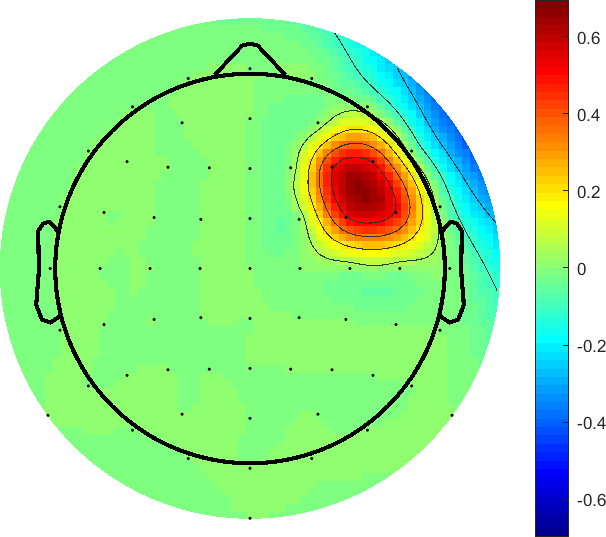}\caption{Alpha}
\end{subfigure}
\caption{Scatter plots for predicted versus true performance measures from LOO-CV MRL models of Table~\ref{BP_table} with the highest adjusted $R^2$. Red and blue distributions in the topographic plots demonstrate the positive and negative weights of significant BP-ROI predictors. ***: $p<0.001$, **: $p<0.01$, *: $p<0.05$.}\label{uni_BPreg_topog}
\end{figure*}

\subsection{Resting-state BP-ROI Feature Relevance Analysis and Regression}\label{pred_results}
As explained in Section~\ref{pred_method}, single linear LOO-CV models were first developed, and based on the statistical significance of the obtained models, all the non-empty subsets of the selected features were used for developing multivariate linear regression models. Table~\ref{BP_table} demonstrates the goodness-of-fits -- obtained from the cross-validated predictions versus true values -- for the best subsets from the entire subset sizes for each of the 8 performance measure-resting state schemes. Figure~\ref{uni_BPreg_topog} demonstrates scatter plots of the true and predicted values for these best models. The significant predictors of each model are also shown in the topographic plots. It is clearly observed how this feature relevance analysis relies on small number of features and can result in highly significant prediction models.

\subsubsection{Pre-task Predictors of Average CVS}
Reduction in the ratio of right parietal alpha, increase in mid-beta (20-24 Hz) from midline parieto-occipital region, and reduction in ratio of upper gamma from left temporal are the best multivariate predictors of better CVSmean during the eyes-open recordings. During the eyes-closed state, reduction of midline parieto-occipital theta and left temporal gamma (28-32 Hz) result in comparable RMSEs in the range of 0.021 to 0.029, $p<0.001$, for both models. These significant predictors appear with similar signs in plots (a) and (b) of NN averaged weights in Figure~\ref{uni_BP_MLPheatmaps}. However, positive weights of the left temporal delta, frontal theta, or midline parieto-occipital gamma observed in the same plots are not entered as the most significant predictors of MVPA models.

\subsubsection{Pre-task Predictors of CVS Variability}
Increase in the ratio of upper beta (24-28 Hz) from right parietal during EO is the single best predictor of increased CVS variability with an adjusted $R^2$ of 0.254, $p<0.01$. During the EC recordings, higher ratios of upper beta and less delta from the left temporal cortex result in a more significant prediction of CVS variability, adj. $R^2$ = 0.574 and RMSE = 0.038, $p<0.001$. The same signs of these significant predictors are observed in plots (c) and (d) in Figure~\ref{uni_BP_MLPheatmaps}, but the frontal alpha or left temporal gamma do not appear in the most significant MVPA models.

\subsubsection{Pre-task Predictors of Average Response Time}
Reduction in the left temporal and right/midline central delta, and increase in the midline parietal delta, right parietal alpha, and left central upper gamma during EO predict slower responses with an extremely small error of 0.392 ms, $p<0.001$. But the midline central and parietal delta had low weights in the plot (e) of Figure~\ref{uni_BP_MLPheatmaps}. During EC, reduction in the right temporal and midline parieto-occipital upper beta and increase in left central upper gamma are significant predictors of slower responses with an error of 37.77 ms, $p<0.001$. 

\subsubsection{Pre-task Predictors of Response Time Variability}
Increase in the ratios of alpha oscillations from the frontal hemisphere during EO, and decrease in the left temporal delta and increase in right frontal alpha during EC are the best predictors of more variability in response time with RMSEs of 0.149 and 0.173, $p<0.05$; however, they result in less significant models compared to the previous predictions.


\section{Discussion} \label{Discuss}
\subsection{Novelty of Experiment Design and Its Automated Pipelines}
The current study focuses on investigating the resting-state neural correlates of score, response time, and their variabilities in a long and monotonous experiment. The preprocessing and artifact rejection steps were performed without subjective manipulations in the selection of independent components. The fixed-sequence nature of the SART experiment ensured to create a boring and repetitive environment while the varying ISIs reduced the chance of repetitive clicks playing a role in increasing the performance scores. Our experiment design managed to drive several participants to complete drowsiness while challenging others to demonstrate their superior skills in maintaining consistent performance and reaction time.  The majority of previous Go/NoGo studies that analyzed the correlates of sustained attention were administered for relatively short intervals and even then, they did not develop regression models for predicting the variability of vigilance scores. Due to the small size of this high-dimensional feature set, shallow neural networks were trained and cross-validated using BP-ROI features to better visualize the polarity and ranking of their hidden unit weights in the obtained multivariate linear regression models, and feature relevance analysis was performed to obtain the most concise and powerful subset of all the 14$\times$12 BP-ROI features from each of the eyes-open and closed EEG recordings. The developed MLR models resulted in successful predictions of the average CVS and HRT from both the EO and EC features and CVS variability from the EC predictors, $p<0.001$, while the CVS variability from EO and HRT variability from both states were less statistically significant. The cross-validated models were proved reliable for extracting the high frequency BP features especially in these small but high dimensional EEG data sets. Thus, the proposed feature extraction method can be used to model the human attention levels for investigation of EEG power ratios originated from sustained attention and performance variations in BCI applications.


In the rest of this section, we discuss the physiological importance of the obtained results. 

\subsection{Roles of Delta and Theta Ratios}\label{discussDelta}
Higher ratios of EO delta especially from the left frontal and temporal regions predicted faster responses as well as less variability in the reaction time in our long SART experiment. Therefore, they could be considered as the neural correlates of uniform impulsivity and point to the role of delta oscillations in improving the ``Go stimulus-responses'' \cite{Karamacoska2018} and suppression of processing the irrelevant stimuli. 
In a much shorter task of auditory Go/NoGo and after rejection of the extremely fast trials as well as the erroneous and post-error trials, increase in delta (1-3 Hz) during the task with respect to the pre-task EO was correlated with higher omission errors and higher RT variability -- standard deviation, according to their calculations \cite{Karamacoska2018}. Their finding on associations between frontal delta with higher RT variability should not be interpreted as being different from our prediction results as we did not eliminate the extremely fast trials and wanted to account for such impulsivities in the reaction time and cumulative vigilance scores. 

The NN heat maps showed that a stronger theta activation from pre-frontal to central regions is correlated with more consistency in CVS and reaction time and, in the case of EO theta, with better CVSmean. However, none of these correlates appeared in the best LOO-CV models of Figure~\ref{uni_BPreg_topog}. Interestingly, higher pre-frontal and frontal theta and right parietal during EO was positively associated with slower reaction time. Similarly, in a study on teenage ADHD and control groups, higher theta (4-7.5 Hz) power from the left frontal cortex and left and right posterior electrodes were found to be correlated with longer CPT reaction time in the control group \cite{Hermens2005}. Their ADHD group showed positive correlations between the pre-task, EO theta oscillations of left frontal sites with both types of errors in the oddball experiment and the false-negative errors in the CPT. The abnormally high left frontal theta activity in ADHD patients was associated with difficulties in detecting signals -- targets - from the background non-target trials and explained by the smaller white matter and lower metabolism in that region. Such associations were not observed for our participants. 

\subsection{Opposite Roles of Frontal and Parietal Alpha in Performance Mean and Variability}
We observed that increase in the alpha ratios from the frontal and central regions was a correlate of lower CVS and higher variabilities in the CVS and HRT; similarly, in an experiment with visual conjunctive continuous performance task (CCPT-V), the group with better performance had demonstrated lower alpha powers during the resting-state and task performance in midline frontal and central and especially the midline parietal cortex \cite{Behzadnia2017}. 

Our findings reveal there is a close relationship between the impaired visual attention and the long-duration task-induced mental fatigue. In particular, changes in the alpha power in both occipital and parietal regions -- Brodmann’s areas 18, 19, and 37 -- are associated with the mental fatigue and result in longer response time and lower CVSmean; likewise, significant correlations were observed between higher tonic alpha from occipital and parietal areas with longer RTs in a vigilance study on simulated driving\cite{Huang2009}. We also observed that increased parieto-occipital alpha was correlated with more consistent CVS and reaction time. These last findings are in line with those of Dockree \textit{et al.} who observed that, in their fixed-SART paradigm, higher tonic alpha powers around 10 Hz were correlated with higher amplitudes of the late positive (LP) ERP that indexed goal activation, better alertness during fixed-SART, and better response patterns \cite{Dockree2007}. They and Fassbender \textit{et al.} believed that the increased activity or desynchronization of alpha power during mentally challenging tasks, such as the random SART, is a sign of effectively remaining on task and blocking attentional drifts and mind-wandering \cite{Fassbender2004}. Thus, participants who demonstrated clear patterns of \textit{maintaining} their vigilance scores throughout the experiment were able to regulate their parietal alpha powers. 

In addition, in a 2-hour simulated experiment, Zheng and Lu \cite{Zheng2017} observed that, in transition to the drowsy states, alpha and theta significantly increased in the temporal and parietal cortex. It was suggested the participants were using a self-regulation strategy to stay on the task despite its monotonous nature. Finally, Loo \textit{et al.} compared ADHD and control groups during a resting state and a CPT session, and observed that weaker lower alpha (8-10 Hz) in the ADHD group was an important biomarker associated with the ``increased cortical arousal'', demands of attention-oriented tasks, and preparation of the visual cortex while expecting the visual stimuli to occur \cite{Loo2009}. 

\subsection{Opposite Roles of Beta Sub-bands in Predicting Task Consistency}
It was suggested that lower parietal alpha demonstrates attentional demands while temporal and parietal beta activities show more differential hemispheric activities during emotional and cognitive processes, respectively \cite{Ray1985}. The heat maps demonstrate that lower beta-1 band (12-16 Hz) is indeed more similar to the lower frequencies in being associated with slower responses while lower beta-2 and mid-beta bands (16-24 Hz) are correlates of improved and consistent performance as well as faster RT in Go trials. We also observed that higher levels of lower-beta (12-20 Hz) from parieto-occipital channels during the EO recordings were correlated with more omission errors. However, increase in the midline and upper beta (20-28 Hz) from pre-frontal and central cortex are associated with better response inhibition or fewer CEs. Increase in mid-beta (20-24 Hz) from midline parietal cortex was also a common predictors of better CVSmean and higher ratios of upper beta from right temporal and midline parietal channels was a significant predictor of short reaction time.

Our results showed that right parietal and left temporal upper beta (24-28 Hz) are the significant predictors of CVS variability from EO and EC band-power features. The positive association of pre-frontal and frontal lower beta and mid-beta ratios, 16-24 Hz, and parietal lower beta-1, 12 - 16 Hz, from EO recordings with more consistent CVS and HRT were demonstrated by the cross-validated neural networks as depicted in Figure~\ref{uni_BP_MLPheatmaps}, but not detected by the multivariate regression models. Although we did not test our participants in terms of having correlates of ADHD, our findings for increased frontal lower beta-2 -- which did not appear as significant predictors in the regression models-- are in line with the increased parietal beta (13--14 Hz) and frontal beta (17 - 18 Hz) being associated with lower variability in the CPT scores of ADHD participants \cite{Loo2009}. The ADHD was found to be associated with increased cortical activation in the form of decreased theta and increased (lower) beta to compensate for the increased arousal -- and weaker slow alpha-- during the resting states \cite{Loo2009}. However, they did not find significant associations between beta-band powers and response time or its variability.

\subsection{Role of Frontal and Midline Parietal Gamma in Smaller Task Variability}
Few pre-task correlates from upper beta and gamma bands are reported for audio or visual Go/NoGo stimuli selection and fatigue especially from the resting state recordings. To fill this gap, our findings demonstrate the role of these narrow and wide-band features in entering the significant prediction models. Our observation on higher gamma ratios from pre-frontal channels being correlated with fewer CEs and lower variability of CVS and HRT do match the existing literature. Increased fronto-parietal gamma was observed in highly experienced meditators \cite{Lutz2004}. In a study on performance variation of BCI systems for healthy participants, differences in the gamma power of two fronto-parietal networks, obtained from the baseline (pre-trial) gamma log-bandpower in the 70-80 Hz range from non-artifactual ICs, predicted within-session variations that occurred within a few minutes in a SMR-BCI classification task \cite{Moritz2012}. The fronto-parietal networks had a positive effect while the pre-frontal and midline parietal sites showed negative coefficients. This prediction was linked with the association of attentional shifts and gamma-range oscillations \cite{Jensen2007}. Our heat maps also show increase in midline parieto-occipital gamma during EO and EC is a predictor of higher CVSmean and lower CVS variability, and faster responses in the EC states.

\subsection{Role of Temporal Gamma in Predicting Task Variability}
On the contrary, increase in the ratios of left central and temporal gamma and upper beta during both EO and EC states were predictors of slower reactions, lower CVSmean, and more CVS variability. Changes in the high frequency EEG oscillations such as gamma and upper beta in predicting behavioral instability and disability to sustain attention can be explained by differences between the high- and low-attention networks at temporal regions (Brodmann’s areas 35 and 36) and the function of default mode network (DMN), a group of brain structures generally thought to be composed of the medial prefrontal cortex, lateral temporal cortex, inferior parietal lobe, and posterior cingulate cortex. It was shown that practicing meditators displayed smaller levels of gamma oscillations in their frontal, central, and temporal regions and higher level of right parieto-occipital gamma with respect to the control participants \cite{Berkovich-Ohana2012}. This network is more active during the wakeful resting states when people are drowning in daydreaming and reflecting on selves or others without any specific reason, and deactivated while they are attending to specific events and tasks. Abnormal DMN activation is also observed in individuals diagnosed with depression, anxiety, and schizophrenia. Similar gamma oscillations related with cognitive processes are also known to be modulated in a number of neural and psychiatric disorders such as the ADHD, bipolar disorder, autism spectrum, and Alzheimer's disease \cite{Fitzgerald2018}. Such impairments can disturb information processing pathways and the synchronization of neuronal clusters. In a study on mindfulness meditation (MM) and DMN, the gamma power in the range of 20 to 45 Hz had decreased with respect to the resting state in the midline and frontal regions in mindfulness practitioners \cite{Berkovich-Ohana2012}. 
The ability to control activity in this network through practicing meditation has been also studied as a way to improve attention and performance in a motor-imagery BCI task as well and resulted in a maximum accuracy of 98\% \cite{Eskandari2008}.

\section{Conclusions}
In this work, we extracted the neural correlates of high tonic vigilance score, fast response time, and consistency of performance and reaction speeds from the pre-task, resting-state EEG signals recorded prior to a 105-minute long SART experiment. We present an automated framework for feature preprocessing and extraction and a novel, adaptive method for vigilance scoring. More importantly, we use neural networks and feature relevance analysis for extraction of the most high ranking and most concise subsets from the band-power ratios of resting intrinsic EEG to predict the vigilance variability and task-related performance measures. The proposed feature extraction method can be used to model the human attention variations for investigation of EEG power ratios originated from sustained attention level in BCI applications. This study will be followed by predicting the long-term correlates of alert and fatigued states using the intrinsic phase synchrony measures and by developing deep neural networks from the extracted features to predict real-time vigilance scores and their variations in individual participants.


\bibliography{JBHI_Resting_SART_BP_arxiv}
\bibliographystyle{ieeetr}
\end{document}